\documentclass[12pt]{article}

\global\arraycolsep=1pt
\usepackage{graphicx}
\usepackage{amssymb}
\usepackage{amsmath}
\usepackage {pstricks}
\usepackage{mathrsfs}

\usepackage{caption} 
\renewcommand{\thefootnote}{\fnsymbol{footnote}}

\numberwithin{equation}{section}
\begin{document}

%
\begin{titlepage}
\begin{flushright}
\texttt{RIKEN-MP-87}\\
\end{flushright}
\vspace{0.5cm}
\begin{center}{\large{\scalebox{1.2}[1.2]{\usefont{T1}{pbk}{b}{sc}On AGT-W Conjecture and }}{\scalebox{1.2}[1.2]{\usefont{T1}{pbk}{b}{it}q}}{\scalebox{1.2}[1.2]{\usefont{T1}{pbk}{b}{sc}-Deformed W-Algebra}}}
\vskip1.0cm
{\large  Masato Taki
}
\vskip 0.8em
{\it
Mathematical Physics Lab., RIKEN Nishina Center,\\ 
Saitama 351-0198, Japan
}%
\\
\vspace{0.3cm}
{\tt taki@riken.jp}
\end{center}
\vskip1.0cm

\begin{abstract}
We propose an extension of the  Alday-Gaiotto-Tachikawa-Wyllard conjecture to 5d $SU(N)$ gauge theories.
A Nekrasov partition function then coincides with the scalar product of the corresponding
Gaiotto-Whittaker vectors of the $q$-deformed $W_N$ algebra.
\end{abstract}
\end{titlepage}


\renewcommand{\thefootnote}{\arabic{footnote}} \setcounter{footnote}{0}


\section{Introduction}

The discovery of the notion of the four-dimensional (4d) $\mathcal{N}=2$ theories of class $\mathcal{S}$ \cite{Gaiotto}
added a new dimension to the study of supersymmetric gauge theories.
The AGT correspondence \cite{AGT} generalized by Wyllard \cite{Wyllard}
is one of the remarkable developments inspired by the study on the 4d theories of class $\mathcal{S}$.
This conjectural correspondence 
relates the 4d theories of class $\mathcal{S}$ with certain 2d conformal field theories (CFTs).
The regular punctures appearing in the M-theory construction \cite{Gaiotto}
are then translated into the corresponding primary states
in the 2d side \cite{AGT,Wyllard},
and the Nekrasov partition function of a 4d theory coincides with
the 2d conformal block for these primary fields.

The AGT-W correspondence was extended by \cite{GaiottoAF,TakiW3}
to the asymptotically-free version of the theories of class $\mathcal{S}$.
This extended class of theories 
arises from M5-branes wrapping a Riemann surface with irregular singularities 
\cite{Marshakov:2009gn,Keller:2011ek,Felinska:2011tn,Bonelli:2011aa,Gaiotto:2012sf,Kanno:2012xt,Nishinaka:2012kn,Kanno:2013vi},
and these singularities lead to Gaiotto-Whittaker vectors in the 2d CFT side.

Recently, 5d uplifts of these 4d $\mathcal{N}=2$ theories have been discussed extensively 
from various view points:
the localization \cite{Hosomichi,Kim:2012gu},
mysterious 6d $(2,0)$ SCFT \cite{Kim:2012ava,Kallen:2012zn,Kim:2012qf},
ultraviolet fixed point theories in 5d \cite{Kim:2012gu,Bergman:2013ala,Taki:2013vka,Bergman:2013aca},
and topological string theory \cite{Awata:2005fa,Iqbal:2007ii,Taki:2007dh,Bao:2011rc,Iqbal:2012xm,Bao:2013pwa,Hayashi:2013qwa,Taki:2014pba}.
The AGT correspondence also has a 5d generalization \cite{AY2009,AY2010}.
The 2d counterpart in the 5d AGT correspondence is
$q$-Virasoro algebra that was introduced in \cite{Shiraishi96,Frenkel,Lukyanov,Feigin96,Awata96} as a 
hidden symmetry of integrable models coming from massive deformation of 2d Virasoro CFTs.
This algebra is generated by $T_n$ that satisfies the following commutation relation
\begin{align}
\left[\,T_n,\,T_m\right]=
-\sum_{\ell=1}^\infty
f_\ell\left(
T_{n-\ell}T_{m+\ell}-T_{m-\ell}T_{n+\ell}
\right)-\frac{(1-q)(1-t^{-1})}{1-p}(p^n-p^{-n})\delta_{n+m,0},
\end{align}
where $p=q/t$.
Awata and Yamada \cite{AY2009} introduced the following Gaiotto-Whittaker vector
\begin{align}
T_1\vert \lambda\rangle
=\lambda\vert \lambda\rangle,\quad
T_{n\geq 2}\vert \lambda\rangle=0,
\end{align}
and they claimed the scalar product of the vector $\vert \lambda=\Lambda^2\rangle$ is equal to the Nekrasov partition function
of 5d $\mathcal{N}=1$ $SU(2)$ super Yang-Mills theory
\begin{align}
Z^{\,SU(2)}_{\,\,\textrm{YM}}=\langle \Lambda^2\vert \Lambda^2\rangle.
\end{align}
In this paper we provide the higher-rank generalization of their finding.
Our conjecture is that the Nekrasov partition function for the $SU(N)$ Yang-Mills theory
is precisely equal to a scalar product of certain Gaiotto-Whittaker vectors of  the $q$-deformed $W_N$ algebra
\begin{align}
\label{conjecture}
Z^{\,SU(N)}_{\,\,\textrm{YM}}=
\langle 0,\cdots,0,\Lambda^N
\vert \Lambda^N,0,\cdots,0\rangle.
\end{align}
To verify this conjecture,
we will present tests of the conjecture at one and two instanton for $N=3$
and one instanton for $N=4$.

This paper is organized as follows. 
We give a brief review on the $q$-deformation of $W$-algebra in section 2. 
In section 3, we identify the Whittaker vectors for the $q$-deformed $W$-algebras that give the Nekrasov partition functions of 5d gauge theories.
We also present explicit checks of this generalized AGT-W correspondence. 
We conclude in section 4. 
In appendix A, we set some conventions for the Nektasov formulas.
Kac-Shapovalov matrixes of $q$-deformed $W_{3}$ and $W_{4}$ algebras are given in appendix B and C.

\section{$\boldsymbol{q}$-deformed $\boldsymbol{W_N}$ algebra}

In this paper, we show that the 5d Nekrasov partition functions are controlled by the hidden 
$q$-deformed $W_N$ symmetry through the AGT-W correspondence.
To understand 2d off-critical integrable models by generalizing the conformal symmetries,
the $q$-deformation of the Virasoro algebra \cite{Shiraishi96,Frenkel,Lukyanov} and $W_N$ algebra \cite{Feigin96,Awata96} were originally introduced.
In this section, we collect some known results on the $q$-deformed $W_N$ algebra,
and we use these formulas in the next section to establish the 5d version of the AGT-W correspondence.

\subsection{$\boldsymbol{q}$-$\boldsymbol{W_3}$ algebra}

Let us  start with writing down the explicitly form of the $q$-$W_3$ algebra.
The $q$-$W_3$ algebra is generated by
two currents $W^1(z)$ and $W^2(z)$,
which are introduced by $q$-deformed version of the Miura transformation \cite{Awata96,Feigin96}.
They satisfy the following relations
\begin{align}
&f^{11}\left(\frac{w}{z}\right)W^1(z)W^1(w)
-W^1(w)W^1(z)f^{11}\left(\frac{z}{w}\right)\nonumber\\
&\qquad\qquad=-\frac{(1-q)(1-t^{-1})}{1-p}
\left(\,\delta\left(\frac{pw}{z}\right)W^2(\sqrt{p}w)
-\delta\left(\frac{w}{pz}\right)W^2\left(\frac{w}{\sqrt{p}}\right)
\right),
\\
&f^{12}\left(\frac{w}{z}\right)W^1(z)W^2(w)
-W^2(w)W^1(z)f^{21}\left(\frac{z}{w}\right)\nonumber\\
&\qquad\qquad=-\frac{(1-q)(1-t^{-1})}{1-p}
\left(\,\delta\left(\frac{p\sqrt{p}w}{z}\right)
-\delta\left(\frac{w}{p\sqrt{p}z}\right)
\right),\\
&f^{22}\left(\frac{w}{z}\right)W^2(z)W^2(w)
-W^2(w)W^2(z)f^{22}\left(\frac{z}{w}\right)\nonumber\\
&\qquad\qquad=-\frac{(1-q)(1-t^{-1})}{1-p}
\left(\,\delta\left(\frac{pw}{z}\right)W^1\left(\frac{z}{\sqrt{p}}\right)
-\delta\left(\frac{w}{pz}\right)W^2\left({\sqrt{p}}w\right)
\right),
\end{align}
where the structure function $f^{\ell m}(z)$ is defined by
\begin{align}
&f^{11}\left({z}\right)=f^{22}\left({z}\right)
=\exp \sum_{n=1}^\infty
\frac{z^n}{n}\frac{(1-p^{2n})(1-q^n)(1-t^{-n})}{1-p^{3n}}
\equiv
\sum_{\ell=0}^\infty f^{11}_\ell z^\ell 
,\\
&f^{12}\left({z}\right)=f^{21}\left({z}\right)
=\exp \sum_{n=1}^\infty
\frac{z^n}{n}\frac{p^{\frac{n}{2}}(1-p^{n})(1-q^n)(1-t^{-n})}{1-p^{3n}}
\equiv
\sum_{\ell=0}^\infty f^{12}_\ell z^\ell .
\end{align}
Notice that the delta function in this article is
$
\delta(z)=\sum_{n\in\mathbb{Z}}z^n
$.

Let us introduce the generators of this algebra.
There are two types of generators $W^{1,2}_n$
because we have two currents $W^{1,2}(z)$.
The mode expansion of the currents is 
\begin{align}
W^1(z)=\sum_{n\in\mathbb{Z}}z^{-n}W^1_n,\quad
W^2(z)=\sum_{n\in\mathbb{Z}}z^{-n}W^2_n.
\end{align}
In this paper we follow the convention of \cite{Awata96}.
The above relations are then equivalent to the following
commutation relations between the generators
\begin{align}
&\left[\,W^1_n,\,W^1_m\right]=
-\sum_{\ell=1}^\infty
f^{11}_\ell\left(
W^1_{n-\ell}W^1_{m+\ell}-W^1_{m-\ell}W^1_{n+\ell}
\right)+d_{n-m}
W^2_{n+m},
\\
&\left[\,W^1_n,\,W^2_m\right]=
-\sum_{\ell=1}^\infty
f^{12}_\ell\left(
W^1_{n-\ell}W^2_{m+\ell}-W^2_{m-\ell}W^1_{n+\ell}
\right)+d_{3n}
\delta_{n+m,0},\\
&\left[\,W^2_n,\,W^2_m\right]=
-\sum_{\ell=1}^\infty
f^{11}_\ell\left(
W^2_{n-\ell}W^2_{m+\ell}-W^2_{m-\ell}W^2_{n+\ell}
\right)+d_{n-m}
W^1_{n+m},
\end{align}
where we introduced
\begin{align}
d_n=c\,(p^{\frac{n}{2}}-p^{-\frac{n}{2}}),
\quad
c=-\frac{(1-q)(1-t^{-1})}{1-p},
\end{align}
for simplicity.
This algebra is not Lie algebra, but we can construct the representation space by considering the Verma module
as the case of the usual Virasoro algebra.
The highest weight vector $\vert \boldsymbol{w} \rangle$ for this algebra is then defined by
\begin{align}
W^\alpha_{0}\vert \boldsymbol{w} \rangle
=w_\alpha\vert \boldsymbol{w} \rangle,\quad
W^\alpha_{n\geq1}\vert \boldsymbol{w} \rangle=0,
\quad \alpha=1,2,
\end{align}
and the Verma module of the algebra is spanned by the vectors $W^1_{n_1}W^1_{n_2}\cdots W^2_{m_1}W^2_{m_2}\cdots\vert \boldsymbol{w} \rangle$
for negative integers $n_i,\,m_j$.
We call the non-negative number $-(\sum_in_i+\sum_j m_j)$ the level of this vector.

\subsection{$\boldsymbol{q}$-$\boldsymbol{W_N}$ algebra and $\boldsymbol{q}$-Miura transformation}

The $q$-deformed $W_N$ algebra is given by the $q$-deformation of the Miura transformation introduced in \cite{Feigin96,Awata96}
\begin{align}
{\boldsymbol{:}}\,
(p^D-A_1(z))
\cdots (p^D-&A_N(p^{1-N}z))\,
{\boldsymbol{:}}
\label{qMiura}
=\sum_{\alpha=1}^N(-1)^\alpha\,W^\alpha(p^{\frac{1-\alpha}{2}}z)\,p^{(N-\alpha)D},
\end{align}
where $p^D$ is the $p$-shift operator since $D\equiv z\partial_z$,
and $A_\alpha(z)$ is the following exponentiated $q$-deformed bosons
\begin{align}
A_\alpha(z)={\boldsymbol{:}}\,\exp\left(
\sum_{n=1}a^\alpha_nz^{-n}
\right)\,
{\boldsymbol{:}}\,
q^{\sqrt{b}\,a^\alpha_0}p^{\frac{N+1}{2}-\alpha}.
\end{align}
This construction is a $q$-deformed version of the free field realization.
The $q$-boson $a_\alpha^n$ is
\begin{align}
[a_\alpha^n, a_\beta^m]=\frac{1}{n}
(1-q^n)\,(1-t^{-n})\frac{1-p^{n(\delta_{\alpha\beta}N-1)}}{1-p^{nN}}p^{nN\theta(\alpha-\beta)}\,\delta_{m+n,0}.
\end{align}
The theta function is defined by $\theta(x<0)=1$ and $\theta(x\geq 0)=0$.
The quantum Miura transformation (\ref{qMiura}) yields the $N-1$ currents
$
W^\alpha(z)=\sum_{n\in\mathbb{Z}} z^{-n}\,W^\alpha_n
$.
Notice that there are only these $N-1$ non-trivial currents because $W^0(z)=W^{ N}(z)=1$ and $W^{\alpha<0}(z)=W^{\alpha> N}(z)=0$.

Using the $q$-boson construction of the currents,
we can show that
the algebraic relation between the $q$-$W$ currents takes the following quadratic form \cite{Odake:2001}
\begin{align}
&f^{\alpha\beta}\left(\frac{w}{z} \right)W^\alpha(z)\,W^\beta(w)
-W^\beta(w)\,W^\alpha(z)\,f^{\beta\alpha}\left(\frac{z}{w} \right)
=c\sum_{\gamma=1}^\alpha\left(
\prod_{\rho=1}^{\gamma-1}
\frac{(1-qp^\rho)(1-t^{-1}p^\rho)}
{(1-p^\rho)(1-p^{\rho+1})}
\right)\nonumber\\
&\times\bigg(\,
\delta\left(p^{\frac{\beta-\alpha}{2}+\gamma}\frac{w}{z}
\right)
f^{\alpha-\gamma\, \beta-\gamma}\left(p^{-\frac{\beta-\alpha}{2}}\right)
W^{\alpha-\gamma}\left(p^{-\frac{\gamma}{2}}z\right)W^{\beta+\gamma}\left(p^{\frac{\gamma}{2}}w\right)\nonumber\\
&\qquad\qquad\qquad-
\delta\left(p^{-\frac{\beta-\alpha}{2}-\gamma}\frac{w}{z}
\right)
f^{\alpha-\gamma\, \beta-\gamma}\left(p^{\frac{\beta-\alpha}{2}}\right)
W^{\alpha-\gamma}\left(p^{\frac{\gamma}{2}}z\right)W^{\beta+\gamma}\left(p^{-\frac{\gamma}{2}}w\right)\bigg)
,
\end{align}
here we adopt the convention $\prod_{\rho=1}^{1-1}*=1$.
The structure functions are given by
\begin{align}
&f^{\alpha\beta}\left(z \right)\equiv
\exp\left(
\sum_{n=1}^\infty
\frac{z^n}{n}(1-q^n)(1-t^{-n})
\frac{1-p^{\alpha n}}{1-p^n}
\frac{1-p^{(N-\beta) n}}{1-p^{Nn}}p^{\frac{\beta-\alpha}{2}n}
\right),\\
&f^{\beta\alpha}\left(z \right)\equiv f^{\alpha\beta}\left(z \right),
\quad(\alpha\leq \beta),
\end{align}
and the Taylor expansion of the structure function around $z=0$ gives the structure constants
$
f^{\alpha\beta}\left(z \right)=\sum_{\ell=0}^\infty
\, f^{\alpha\beta}_\ell\,z^\ell
$.
The highest weight vector $\vert \boldsymbol{w} \rangle$ for this algebra is then defined by
\begin{align}
W^\alpha_{0}\vert \boldsymbol{w} \rangle
=w_\alpha\vert \boldsymbol{w} \rangle,\quad
W^\alpha_{n\geq1}\vert \boldsymbol{w} \rangle=0,
\quad \alpha=1,2,\cdots, N-1.
\end{align}
We can also construct the Verma module by acting lowering operators $W^\alpha_{n<0}$ succesively.

For the latter discussion, let us write down the commutation relations in the $N=4$ case explicitly.
There are three currents, and their generators satisfy 
\begin{align}
&\left[\,W^1_n,\,W^1_m\right]=
-\sum_{\ell=1}^\infty
f^{11}_\ell\left(
W^1_{n-\ell}W^1_{m+\ell}-W^1_{m-\ell}W^1_{n+\ell}
\right)+d_{n-m}
W^2_{n+m},
\\
&\left[\,W^2_n,\,W^2_m\right]=
-\sum_{\ell=1}^\infty
f^{22}_\ell\left(
W^2_{n-\ell}W^2_{m+\ell}-W^2_{m-\ell}W^2_{n+\ell}
\right)+\frac{(1-qp)(1-t^{-1}p)}{(1-p)(1-p^2)}d_{4n}\delta_{n+m}
\nonumber\\
&\quad\,\,
+c\,d_{2n}\frac{1+p^2}{1-p^2}
\delta_{n+m}
+d_{n-m}\sum_{r=0}^\infty\left(\sum_{\ell=0}^r f^{13}_\ell\right)\left(
W^1_{-r}W^3_{n+m+r}+W^3_{n+m-r-1}W^1_{r+1}
\right)
,
\\
&\left[\,W^3_n,\,W^3_m\right]=
-\sum_{\ell=1}^\infty
f^{33}_\ell\left(
W^3_{n-\ell}W^3_{m+\ell}-W^3_{m-\ell}W^3_{n+\ell}
\right)+d_{n-m}
W^2_{n+m},
\\
&\left[\,W^1_n,\,W^2_m\right]=
-\sum_{\ell=1}^\infty
f^{12}_\ell\left(
W^1_{n-\ell}W^2_{m+\ell}-W^2_{m-\ell}W^1_{n+\ell}
\right)+d_{2n-m}
W^3_{n+m,0},
\\
&\left[\,W^1_n,\,W^3_m\right]=
-\sum_{\ell=1}^\infty
f^{13}_\ell\left(
W^1_{n-\ell}W^3_{m+\ell}-W^3_{m-\ell}W^1_{n+\ell}
\right)+d_{4n}
\delta_{n+m,0},\\
&\left[\,W^2_n,\,W^3_m\right]=
-\sum_{\ell=1}^\infty
f^{23}_\ell\left(
W^2_{n-\ell}W^3_{m+\ell}-W^3_{m-\ell}W^2_{n+\ell}
\right)+d_{n-2m}
W^1_{n+m,0}.
\end{align}
The Kac-Shapovalov matrix at level one is given in Appendix.C.

Notice that there are no algebraic distinction between $W^1$ and $W^3$.
In the generic $q$-$W_N$ case, the reflection $W^\alpha\leftrightarrow W^{N-\alpha}$ is a symmetry.
This fact is a key to specify the explicit form (\ref{conjecture}) of the 5d generalization of the AGT-W correspondence.

\section{The AGT-W correspondence in 5d}

In this section,
we demonstrate that 5d Nekrasov instanton partition functions
are precisely equal to the scalar norms of the corresponding Whittaker vectors for $q$-$W$ algebra.
This result provides 5d generalization of the AGT-W correspondence for irregular singularities.
The basic element in the gauge theory side of this correspondence is 
the Nekrasov instanton partition function \cite{Nekrasov:2002qd}.
This partition function for 5d $SU(N)$ Yang-Mills theory takes the following form
\begin{align}
\label{SUNNek}
Z^{\,SU(N)}_{\,\,\textrm{YM}}=
\sum_{\vec{Y}}\,
\left(
\Lambda^2
\sqrt{\frac{q}{t}}
\right)^{{N\vert\vec{Y}\vert}}
\prod_{\alpha,\beta=1}^N
\frac{1}{N_{Y_\alpha Y_\beta}
(Q_\beta Q_\alpha^{-1})
}
=\sum_{k=0}^\infty\,
\Lambda^{2Nk}\,
Z^{\,SU(N)}_{\,\,\textrm{YM}\,k\textrm{-inst.}},
\end{align}
where $\vec{Y}$ is $N$-tuple of Young diagrams.
The parameter $Q_\alpha=e^{-Ra_\alpha}$ is an exponentiated Coulomb branch parameter,
and therefore it satisfies $Q_1Q_2\cdots Q_N=1$.
Using this formula and summing over Young diagrams, we can easily calculate this partition function.
In the remaining part of this section, we consider $q$-$W_N$ algebra description of the same result.

\subsection{$\boldsymbol{SU(3)}$ Yang-Mills versus $\boldsymbol{q}$-$\boldsymbol{W_3}$ algebra}

A Whittaker vector is a state in the Verma module of a given algebra, 
and this vector is characterized by certain coherent state conditions for the lowering operators.
The lowering operators for the $q$-$W_3$ algebra are the generators with negative levels $W^{1,2}_{n<0}$.
We then introduce the following Whittaker vector of the $q$-$W_3$ algebra 
\begin{align}
\label{W3pure-1}
&W^1_1\vert \Lambda_1,\Lambda_2\rangle
=\Lambda_1\vert \Lambda_1,\Lambda_2\rangle,\\
&W^2_1\vert \Lambda_1,\Lambda_2\rangle
=\Lambda_2\vert \Lambda_1,\Lambda_2\rangle,\\
\label{W3pure-3}
&W^1_n\vert \Lambda_1,\Lambda_2\rangle
=W^2_n\vert \Lambda_1,\Lambda_2\rangle=0,\quad n\geq2.
\end{align}
We can impose this condition consistently with the commutation relations.
This Whittaker vector is the $q$-deformed version of that for the usual $W_3$ algebra \cite{TakiW3,Kanno:2012xt}.
As a vector in the Verma module,
this Whitaker state is determined 
by the above defining equations up to overall factor.
We will normalize the state as $\vert \Lambda_1,\Lambda_2\rangle
=\vert \boldsymbol{w} \rangle+\cdots$.
We parametrize the highest weights $\boldsymbol{w}$ as follows
\begin{align}
\label{qW3weights}
w_1=Q_1+Q_2+\frac{1}{Q_1Q_2},\quad
w_2=\frac{1}{Q_1}+\frac{1}{Q_2}+Q_1Q_2.
\end{align}
Our conjecture is that the instanton partition function of
5d $SU(3)$ pure super Yang-Mills theory
is equal to the scalar product of two Whittaker vectors of the $q$-$W_3$ algebra:
\begin{align}
\label{conj3}
Z^{\,SU(3)}_{\,\,\textrm{YM}}=
\langle \,0,\Lambda^3\,
\vert \,\Lambda^3,0\,\rangle=
\langle \,\Lambda^3,0\,
\vert \,0,\Lambda^3\,\rangle.
\end{align}
Here $\langle \,0,\Lambda^3\,
\vert$ is the dual vector of $\vert \,0,\Lambda^3\,\rangle$.
This conjecture is precisely 5d generalization of the AGT-W correspondence
between 4d $SU(3)$ pure Yang-Mills theory 
and the $W_3$ algebra proposed in \cite{TakiW3}. 
We give an explicit check of this 5d relation (\ref{conj3}) in the following.

\subsubsection*{one-instanton test}

It is easy to compute the one instanton partition function by using the formula (\ref{SUNNek}).
For instance, the contribution from the 3-tuple $\vec{Y}=([1],\emptyset,\emptyset)$
is
\begin{align}
\left(\sqrt{\frac{q}{t}}\right)^3
\frac{1}{(1-q)(1-t^{-1})(1-Q_2Q_1^{-1}t^{-1}q)
(1-Q_3Q_1^{-1}t^{-1}q)
(1-Q_1Q_2^{-1})(1-Q_1Q_3^{-1})
}.
\end{align}
The 1-instanton partition function is the summation of three contributions
coming from the 3-tuples satisfying $\vert \vec{Y}\vert=1$.
Eliminating $Q_3$ by the constraint $Q_3=(Q_1Q_2)^{-1}$ yields the following result
\begin{align}
\label{SU3pure1inst}
&Z^{\,SU(3)}_{\,\,\textrm{YM}\,1\textrm{-inst.}}
=-q^{\frac{3}{2}}t^{\frac{3}{2}}Q_1^2Q_2^2
\nonumber\\
&\times
\frac{
q^2t^2\left(Q_1^3Q_2^4+Q_1^4Q_2^3
+Q_1Q_2^3+Q_1^3Q_2
+Q_1+Q_2\right)
-\left(t^4
+2qt^3+2q^3t+q^4
\right)Q_1^2Q_2^2
}{(1-q)(1-t)
(qQ_1-tQ_2)
(tQ_1-qQ_2)
(qQ_1^2Q_2-t)
(tQ_1^2Q_2-q)
(qQ_1Q_2^2-t)
(tQ_1Q_2^2-q)
}.
\end{align}
Let us reproduce this result by means of the Whittaker vectors of the $q$-$W_3$ algebra.

Since the Whittaker vector lies in the Verma module,
we can use the ansatz
\begin{align}
\vert \Lambda_1,\Lambda_2\rangle
=\vert \boldsymbol{w} \rangle
+c_{10}W^1_{-1}\vert \boldsymbol{w} \rangle
+c_{01}W^2_{-1}\vert \boldsymbol{w} \rangle+\cdots.
\end{align}
Imposing the conditions (\ref{W3pure-1}-\ref{W3pure-3}) at  level one,
we can solve it at this level by using the level one Kac-Shapovalov matrix $\mathcal{G}^{(1)}$
that is the following Gram matrix
\begin{align}
 \left(\begin{tabular}{c} $\Lambda_1$\\ $\Lambda_2$\end{tabular}\right)
 =\mathcal{G}^{(1)}\,
  \left(\begin{tabular}{c} $c_{10}$\\ $c_{01}$\end{tabular}\right),
\quad
\mathcal{G}^{(1)}=\left(\begin{tabular}{cc}  
$\langle{\boldsymbol{w}}\vert W^1_{1}W^1_{-1}\vert{\boldsymbol{w}}\rangle$
&
$\langle{\boldsymbol{w}}\vert W^1_{1}W^2_{-1}\vert{\boldsymbol{w}}\rangle$
\\
$\langle{\boldsymbol{w}}\vert W^2_{1}W^1_{-1}\vert{\boldsymbol{w}}\rangle$
& 
$\langle{\boldsymbol{w}}\vert W^2_{1}W^2_{-1}\vert{\boldsymbol{w}}\rangle$
\end{tabular}\right).
\end{align}
The components are given by
\begin{align}
\langle{\boldsymbol{w}}\vert W^1_{1}W^1_{-1}\vert{\boldsymbol{w}}\rangle&
=-f^{11}_1(w_1)^2
+d_2
w_2,
\\
\langle{\boldsymbol{w}}\vert W^1_{1}W^2_{-1}\vert{\boldsymbol{w}}\rangle&
=\langle{\boldsymbol{w}}\vert W^1_{2}W^1_{-1}\vert{\boldsymbol{w}}\rangle
=-f^{12}_1w_1w_2
+d_3,
\\
\langle{\boldsymbol{w}}\vert W^2_{1}W^2_{-1}\vert{\boldsymbol{w}}\rangle&
=-f^{11}_1(w_2)^2
+d_2
w_1.
\end{align}
The solution to the Whittaker condition (\ref{W3pure-1}-\ref{W3pure-3}) at level one
then leads to
\begin{align}
\langle \Lambda'_1,\Lambda'_2
\vert \Lambda_1,\Lambda_2\rangle
=1
+
 \left(\begin{tabular}{cc} $\Lambda'_1$& $\Lambda'_2$\end{tabular}\right)\,\left(\mathcal{G}^{(1)}\right)^{-1}\,
  \left(\begin{tabular}{c} $\Lambda_1$\\ $\Lambda_2$\end{tabular}\right)
+\cdots.
\end{align}
Notice that $\langle \Lambda_1,\Lambda_2
\vert W^\alpha_{-1}=\langle \Lambda_1,\Lambda_2
\vert \Lambda_\alpha$.
We can also solve the condition (\ref{W3pure-1}-\ref{W3pure-3}) at a given level in this way.
The level $k$ contribution is then given by 
the level $k$ Kac-Shapovalov matrix $\mathcal{G}^{(k)}$ for the $q$-$W_3$ algebra.
Consider the following specialized situation
\begin{align}
\label{normWG3}
\langle 0,\Lambda^3
\vert \Lambda^3,0\rangle
=1
+
\Lambda^6\,\left(\mathcal{G}^{(1)}\right)^{-1}_{21}
+\cdots
=1+\sum_{k=1}^\infty
\Lambda^{6k}\,\left(\mathcal{G}^{(k)}\right)^{-1}_{{}^{{(W^2_{-1})^k,(W^1_{-1})^k}}}.
\end{align}
In this case,
the level $k$ contribution reduces to the $(\,(W^2_{-1})^k,(W^1_{-1})^k\,)$-component.
Then our conjecture (\ref{conj3}) at level one claims that $\left(\mathcal{G}^{(1)}\right)^{-1}_{21}$ coincides with the one instanton
partition function of 5d $SU(3)$ pure Yang-Mills theory.
Substituting our parametrization (\ref{qW3weights}),
the off-diagonal component of the matrix $\left(\mathcal{G}^{(1)}\right)^{-1}$ takes the following form
\begin{align}
&
\left(\mathcal{G}^{(1)}\right)^{-1}_{21}=
\frac{-\langle{\boldsymbol{w}}\vert W^2_{1}W^1_{-1}\vert{\boldsymbol{w}}\rangle}
{\det \mathcal{G}^{(1)}}=-q^{\frac{3}{2}}t^{\frac{3}{2}}Q_1^2Q_2^2
\nonumber\\
&\times
\frac{
q^2t^2\left(Q_1^3Q_2^4+Q_1^4Q_2^3
+Q_1Q_2^3+Q_1^3Q_2
+Q_1+Q_2\right)
-\left(t^4
+2qt^3+2q^3t+q^4
\right)Q_1^2Q_2^2
}{(1-q)(1-t)
(qQ_1-tQ_2)
(tQ_1-qQ_2)
(qQ_1^2Q_2-t)
(tQ_1^2Q_2-q)
(qQ_1Q_2^2-t)
(tQ_1Q_2^2-q)
}.
\end{align}
This is precisely equal  to (\ref{SU3pure1inst}).

\subsubsection*{two-instanton test}

We can also present explicit check of our conjecture (\ref{conj3}) at two instanton.
At two instanton $\vert \vec{Y}\vert=2$, the partition function
has the contributions from the 3-tuples $\vec{Y}=([1],[1],\emptyset),$ $([2],\emptyset,\emptyset),$ $([1^2],\emptyset,\emptyset)$ and their permutations.
Substituting these 3-tuples of Young diagrams into (\ref{SUNNek})
yields the two instanton partition function $Z^{\,SU(3)}_{\,\,\textrm{YM}\,2\textrm{-inst.}}$ for $SU(3)$ pure Yang-Mills theory.

The ${q}$-${W_3}$ counterpart to the two instanton partition function is the level two Kac-Shapovalov matrix as (\ref{conj3}) and (\ref{normWG3}).
The explicit form of the Kac-Shapovalov matrix $\mathcal{G}^{(2)}$ is given in Appendix.B.
By computer calculation, we can show that the level two part of the norm (\ref{normWG3}) of the Whittaker vectors
is equal to the two instanton partition function as
\begin{align}
Z^{\,SU(3)}_{\,\,\textrm{YM}\,2\textrm{-inst.}}
=\frac{\widetilde{\left(\mathcal{G}^{(2)}\right)}_{25}}{\det \mathcal{G}^{(2)}},
\end{align}
where $\widetilde{\left(\mathcal{G}^{(2)}\right)}_{25}$ is the $(2,5)$-cofactor of the level two Kac-Shapovalov matrix.
This result confirms our conjecture at level two.

\subsection{$\boldsymbol{SU(3)}$ SQCD versus $\boldsymbol{q}$-$\boldsymbol{W_3}$ algebra}

We can generalize our conjecture to $SU(3)$ gauge theory with fundamental hypermultiplets.
Let us consider the $SU(3)$ gauge theory with single fundamental hypermultiplet for example.
The Nekrasov partition function for this theory is
\begin{align}
Z^{\,SU(3)}_{\,\,N_f=1}=
\sum_{\vec{Y}}\,
\Lambda^{5\vert\vec{Y}\vert}
\left(
\sqrt{\frac{q}{t}}
\right)^{{3\vert\vec{Y}\vert}}
\frac{
\prod_{\alpha=1}^3 \prod_{(i,j)\in Y_\alpha}
(1-Q_\alpha Q_m t^{-i+1}q^{j-1})
}{
\prod_{\alpha,\beta=1}^3
N_{Y_\alpha Y_\beta}
(Q_\beta Q_\alpha^{-1})
},
\end{align}
where $Q_m=e^{-Rm}$ is the exponentiated mass parameter.

The Nekrasov partition function of this SQCD
should be equal to the scalar product of the two Whittaker vectors
\begin{align}
\label{conjSU3QCD}
Z^{\,SU(3)}_{\,\,N_f=1}=\langle \,0,\Lambda^3\,
\vert\, \Lambda^2,Q_m\Lambda^2\,\rangle.
\end{align}
It is easy to check this conjecture at one instanton level.
The one instanton part actually coincides with the following combination of elements
of the inverse Kac-Shapovalov matrix at level one
\begin{align}
&Z^{\,SU(3)}_{\,\,N_f=1\,1\textrm{-inst.}}=
Z^{\,SU(3)}_{\,\,\textrm{YM}\,1\textrm{-inst.}}+q^{{2}}t^{{2}}Q_1^2Q_2^2Q_m\times
\nonumber\\
&
\frac{
\left({qt^2}+{q^2t}
\right)\left(Q_1^4Q_2^4
+Q_1^2+Q_2^2
\right)
-\left({t^3}+{q^3}
\right)\left(Q_1^2Q_2^3
+Q_1^3Q_2^2
+Q_1Q_2\right)
}{(1-q)(1-t)
(qQ_1-tQ_2)
(tQ_1-qQ_2)
(qQ_1^2Q_2-t)
(tQ_1^2Q_2-q)
(qQ_1Q_2^2-t)
(tQ_1Q_2^2-q)
}\nonumber\\
&=\left(\mathcal{G}^{(1)}\right)^{-1}_{21}+Q_m\left(\mathcal{G}^{(1)}\right)^{-1}_{22},
\end{align}
and this combination is precisely the level one part of the right hand side go our conjecture (\ref{conjSU3QCD}).
By employing compter calculation, we should be able to justify this conjecture at higher level.
Further generalization of this relation is also straightforward. The scalar product $\langle \,Q_m\Lambda^2,\Lambda^2\,
\vert\, \Lambda^2,Q_m'\Lambda^2\,\rangle$ corresponds to the Nekrasov partition function of $SU(3)$ gauge theory with two fundamental hypermultiplets.

\subsection{generalization to $\boldsymbol{q}$-$\boldsymbol{W_N}$ algebra}

Generalization of our proposal (\ref{conj3}) to the higher rank cases is also very simple.
Our conjecture is that the instanton partition function of
5d $SU(N)$ pure super Yang-Mills theory
is equal to the scalar product of two Whittaker vectors of the $q$-$W_N$ algebra:
\begin{align}
\label{conjN}
Z^{\,SU(N)}_{\,\,\textrm{YM}}=
\langle 0,\cdots,0,\Lambda^N
\vert \Lambda^N,0,\cdots,0\rangle.
\end{align}
The Whittaker vectors are defined by
\begin{align}
&W^\alpha_1\vert \Lambda_1,\cdots, \Lambda_{N-1}\rangle
=\Lambda_\alpha\vert \Lambda_1,\cdots, \Lambda_{N-1}\rangle,
\quad W^\alpha_n\vert \Lambda_1,\cdots, \Lambda_{N-1}\rangle
=0,\quad n\geq2,\\
&\langle\Lambda_1,\cdots, \Lambda_{N-1}\vert  W^\alpha_{-1}
=\Lambda_\alpha\langle \Lambda_1,\cdots, \Lambda_{N-1}\vert ,
\quad \langle \Lambda_1,\cdots, \Lambda_{N-1}\vert W^\alpha_m 
=0,\quad m\leq-2,
\end{align}
and the identification between the Coulomb branch parameters and the highest weights is
\begin{align}
w_\alpha=\sum_{1\leq\beta_1<\cdots<\beta_\alpha\leq N}
Q_{\beta_1}\cdots Q_{\beta_\alpha},\quad \alpha=1,2,\cdots,N-1.
\end{align}
Notice that $Q_1Q_2\cdots Q_N=1$.

Since it is hard to prove this conjecture,
we show the relation at one instanton level in the $SU(4)$ case.
Higher rank generalization of this test also should be straightforward.

\subsubsection*{one-instanton test for $\boldsymbol{q}$-$\boldsymbol{W_4}$ algebra}

Let us test our conjecture (\ref{conjN}) for $N=4$.
The corresponding algebra is $q$-$W_4$ algebra
whose commutation relations were given in the previous section.
The level one Kac-Shapovalov matrix $\mathcal{G}^{(1)}\left(w_1,w_2,w_3,w_4 \right)$ for this algebra is given in Appendix.C.
As we showed for $N=3$,
the level one part of the scalar product in the right hand side of (\ref{conjN}) is given by the $(3,1)$-component of the inverse matrix $(\mathcal{G}^{(1)})^{-1}$.
Using computer calculation, we can easily show the one instanton ({\it i.e.} $\Lambda^8$) part of our conjecture (\ref{conjN}) 
\begin{align}
Z^{\,SU(4)}_{\,\,\textrm{YM}\,1\textrm{-inst.}}
&={\scriptstyle
\frac{q^2t^2\left( q+t\right)\left( Q_1Q_2Q_3\right)^2\left(
q^8Q_1^4Q_2^4Q_3^4
-2q^7tQ_1^4Q_2^4Q_3^4
+q^6t^2\left(Q_1^4Q_2^6Q_3^6+2Q_1^5Q_2^6Q_3^5+\cdots\right)
+\cdots
\right)}
{\left(1-q\right)\left(1-t\right)
\prod_{(\alpha,\beta,\gamma)=
(1,2,3),(2,3,1),(3,1,2)}\left(qQ_\beta-tQ_\alpha\right)\left(tQ_\beta-qQ_\alpha\right)
\left(qQ_\alpha^2Q_\beta Q_\gamma-t\right)\left(tQ_\alpha^2Q_\beta Q_\gamma-q\right)
}
}
\nonumber\\
&=
\left( \mathcal{G}^{(1)}\right)^{-1}_{31}.
\end{align}
To show this relation, we substitute $Q_4=(Q_1Q_2Q_3)^{-1}$.

\section{Discussion}

In this paper we proposed the generalized AGT-W correspondence between
5d uplift of 4d $\mathcal{N}=2$ $SU(N)$ asymptotically-free gauge theories and the $q$-deformed $W_N$ algebra.
The Nekrasov partition function of a 5d gauge theory is then equal to the scalar product of the corresponding Whittaker vectors
of the $q$-deformed $W_N$ algebra.
We presented explicit checks of our conjecture based on the instanton expansion.

Our conjecture simplifies the original 4d AGT-W correspondence \cite{GaiottoAF,TakiW3}
since we can write down the algebraic relations of the $q$-deformed $W_N$ algebra explicitly,
although it is very hard to find the explicit form for the undeformed one.
The complicated structure of the original $W_N$ algebra
is packed into the embedding of the $W_N$ currents in the deformed currents.
The 4d limit, which is easy to see in the gauge theory side,
of our 5d conjecture should shed new light on the 4d AGT-W correspondence \cite{GaiottoAF,TakiW3}.

There are many further directions.
Generalization to 5d $\mathcal{N}=1$ $SU(N)$ SQCD with $N_f
$ flavors should be possible since
we can expect the following relation
\begin{align}
Z^{\,SU(N)}_{\,\,N_f
}=
Z^{\,U(1)}_{\,\,N_f}\cdot
\langle \Lambda'_1,\cdots,\Lambda'_{N-1}
\vert \Lambda_{N-1},\cdots,\Lambda_1\rangle.
\end{align}
It would also interesting if we can generalize our conjecture
to generic ABCDEFG gauge groups and $q$-deformed $W$-algebras.
Such generalization in 4d case was already studied in \cite{Keller:2011ek}.
Proof of the original AGT-W relation are already known \cite{Schiffmann,Okounkov},
and therefore there should be some relation between our conjecture and the quantum groups appearing in \cite{Schiffmann,Okounkov}.
It should also be possible to verify our conjecture along the line of \cite{Alba,Fateev,Awata:2011ce,Yanagida}.

\section*{Acknowledgments}
We are very grateful to Vladimir Mitev, Elli Pomoni and
Futoshi Yagi for fruitful discussions during collaborations.


\appendix
\section{Nekrasov formula}

The Nekrasov partition functions for the 5d theories on $\mathbb{R}^4\times S^1$, which are uplift of the 4d $\mathcal{N}=2$ gauge theories,
is an equivariant index for the instanton moduli space \cite{Nekrasov:2002qd}.
Employing the Duistermaat-Heckman formula, Nekrasov derived a closed expression for the partition functions
based on the Chern characters of the complex for the linearized ADHM equations.
The $SU(N)$ vector multiplet contribution to the Nekrasov partition functions is
\begin{align}
Z^{\,\textrm{vect.}}_{\,\vec{Y}}
(Q_{\alpha};t,q)
=\left(\sqrt{\frac{q}{t}}\right)^{N{\vert\vec{Y}\vert}}
\frac{1}{\prod_{\alpha,\beta=1,2}N_{Y_\alpha Y_\beta}(Q_{\beta\alpha};t,q)},
\end{align}
where $Q_{\alpha\beta}=Q_\alpha Q_\beta^{-1}$, $Q_\alpha=e^{-Ra_\alpha}$ is the exponentiated Coulomb branch parameters
and so they satisfy $\prod_\alpha Q_\alpha=1$.
The factor $N_{Y_\alpha,Y_\beta}$ is given by
\begin{align}
N_{Y_\alpha Y_\beta}(Q;t,q)
&=
\prod_{s\in R_\alpha}
\left( 1-Q\,t^{\ell_{Y_\beta}(s)}q^{a_{Y_\alpha}(s)+1} \right)
\prod_{t\in Y_\beta}
\left( 1-Q\,t^{-(\ell_{Y_\alpha}(t)+1)}q^{-a_{Y_\beta}(t)} \right).
\end{align}
The arm and leg length here is defined by
\begin{align}
a_Y(i,j)=Y_i-j,\quad\ell_Y(i,j)=Y^t_j-i.
\end{align}
The fundamental hypermultiplets lead to the following contribution
\begin{align}
Z^{\,\textrm{fund.}}_{\,\vec{Y}}Q_{\alpha},Q_m;t,q)
=\prod_{\alpha=1}^N \prod_{(i,j)\in Y_\alpha}
(1-Q_\alpha Q_m t^{-i+1}q^{j-1}).
\end{align}
The instanton number $k$ corresponds to the total number of the boxes in the N-tuples $\vec{Y}$ of Young diagrams  as
$k=\vert\vec{Y}\vert:=\sum_\alpha \vert{Y}_\alpha\vert$.
The $k$-instanton partition function for $SU(N)$ gauge theory with $N_f$ flavors is therefore
\begin{align}
Z^{\,SU(N)}_{\,\,\textrm{SQCD}\,k\textrm{-inst.}}(Q_{\alpha},Q_{mf};t,q)=\sum_{\vert\vec{Y}\vert=k}\,
Z^{\,\textrm{vect.}}_{\,\vec{Y}}
(Q_{\alpha};t,q)\prod_{f=1}^{N_f}Z^{\,\textrm{fund.}}_{\,\vec{Y}}(Q_{\alpha},Q_{mf};t,q).
\end{align}

\section{$\boldsymbol{q}$-$\boldsymbol{W_3}$ Kac-Shapovalov matrix at level-two }

The Kac-Shapovalov matrix for ${q}$-deformed ${W_3}$ algebra at level-two
\begin{align}
&\mathcal{G}^{(2)}\left(w_1,w_2,w_3 \right)=\nonumber\\
&\left(\begin{tabular}{ccccc}  
${\scriptstyle \langle{\boldsymbol{w}}\vert W^1_{2}W^1_{-2}\vert{\boldsymbol{w}}\rangle}$
&
${\scriptstyle \langle{\boldsymbol{w}}\vert W^1_{2}(W^1_{-1})^2\vert{\boldsymbol{w}}\rangle}$
&
${\scriptstyle \langle{\boldsymbol{w}}\vert W^1_{2}W^1_{-1}W^2_{-1}\vert{\boldsymbol{w}}\rangle}$
&
${\scriptstyle \langle{\boldsymbol{w}}\vert W^1_{2}W^2_{-2}\vert{\boldsymbol{w}}\rangle}$
&
${\scriptstyle \langle{\boldsymbol{w}}\vert W^1_{2}(W^2_{-1})^2\vert{\boldsymbol{w}}\rangle}$
\\
${\scriptstyle \langle{\boldsymbol{w}}\vert (W^1_{1})^2W^1_{-2}\vert{\boldsymbol{w}}\rangle}$
&
${\scriptstyle \langle{\boldsymbol{w}}\vert (W^1_{1})^2(W^1_{-1})^2\vert{\boldsymbol{w}}\rangle}$
&
${\scriptstyle \langle{\boldsymbol{w}}\vert (W^1_{1})^2W^1_{-1}W^2_{-1}\vert{\boldsymbol{w}}\rangle}$
&
${\scriptstyle \langle{\boldsymbol{w}}\vert (W^1_{1})^2W^2_{-2}\vert{\boldsymbol{w}}\rangle}$
&
${\scriptstyle \langle{\boldsymbol{w}}\vert (W^1_{1})^2(W^2_{-1})^2\vert{\boldsymbol{w}}\rangle}$
\\
${\scriptstyle \langle{\boldsymbol{w}}\vert W^2_{1}W^1_{1}W^1_{-2}\vert{\boldsymbol{w}}\rangle}$
&
${\scriptstyle \langle{\boldsymbol{w}}\vert W^2_{1}W^1_{1}(W^1_{-1})^2\vert{\boldsymbol{w}}\rangle}$
&
${\scriptstyle \langle{\boldsymbol{w}}\vert W^2_{1}W^1_{1}W^1_{-1}W^2_{-1}\vert{\boldsymbol{w}}\rangle}$
&
${\scriptstyle \langle{\boldsymbol{w}}\vert W^2_{1}W^1_{1}W^2_{-2}\vert{\boldsymbol{w}}\rangle}$
&
${\scriptstyle \langle{\boldsymbol{w}}\vert W^2_{1}W^1_{1}(W^2_{-1})^2\vert{\boldsymbol{w}}\rangle}$
\\
${\scriptstyle \langle{\boldsymbol{w}}\vert W^2_{2}W^1_{-2}\vert{\boldsymbol{w}}\rangle}$
&
${\scriptstyle \langle{\boldsymbol{w}}\vert W^2_{2}(W^1_{-1})^2\vert{\boldsymbol{w}}\rangle}$
&
${\scriptstyle \langle{\boldsymbol{w}}\vert W^2_{2}W^1_{-1}W^2_{-1}\vert{\boldsymbol{w}}\rangle}$
&
${\scriptstyle \langle{\boldsymbol{w}}\vert W^2_{2}W^2_{-2}\vert{\boldsymbol{w}}\rangle}$
&
${\scriptstyle \langle{\boldsymbol{w}}\vert W^2_{2}(W^2_{-1})^2\vert{\boldsymbol{w}}\rangle}$
\\
${\scriptstyle \langle{\boldsymbol{w}}\vert (W^2_{1})^2W^1_{-2}\vert{\boldsymbol{w}}\rangle}$
&
${\scriptstyle \langle{\boldsymbol{w}}\vert (W^2_{1})^2(W^1_{-1})^2\vert{\boldsymbol{w}}\rangle}$
&
${\scriptstyle \langle{\boldsymbol{w}}\vert (W^2_{1})^2W^1_{-1}W^2_{-1}\vert{\boldsymbol{w}}\rangle}$
&
${\scriptstyle \langle{\boldsymbol{w}}\vert (W^2_{1})^2W^2_{-2}\vert{\boldsymbol{w}}\rangle}$
&
${\scriptstyle \langle{\boldsymbol{w}}\vert (W^2_{1})^2(W^2_{-1})^2\vert{\boldsymbol{w}}\rangle}$
\end{tabular}\right),\nonumber
\end{align}
is a symmetric matrix,
whose matrix elements are given by
\begin{align}
&\mathcal{G}_{11}^{(2)}=-f^{11}_1\,\mathcal{G}_{11}^{(1)}-f^{11}_2\,(w_1)^2+d_4\,w_2
,\\
&\mathcal{G}_{22}^{(2)}=\left(1-f^{11}_2\right)\left(\mathcal{G}_{11}^{(1)}\right)^2
+\left(
-f^{11}_1\left(1-f^{11}_1\right)^2\,(w_1)^2+d_2\,w_2
\right)\mathcal{G}_{11}^{(1)}\nonumber\\
&\qquad\quad-
\left(
2f^{11}_1\left(1-f^{11}_1\right)\,d_1\,w_1+f^{12}_1d_2\,w_1
\right)\mathcal{G}_{12}^{(1)}
-f^{11}_1\,(d_1)^2\mathcal{G}_{22}^{(1)}
,\\
&\mathcal{G}_{33}^{(2)}=\left(1-f^{11}_2\right)\left(\mathcal{G}_{12}^{(1)}\right)^2
-f^{11}_1\left(f^{12}_1\right)^2(w_2)^2\,\mathcal{G}_{11}^{(1)}
\nonumber\\
&\qquad\quad
+\left(
2f^{11}_1f^{12}_1w_1w_2+d_1d_2
\right)\mathcal{G}_{12}^{(1)}+
\left(
\left(1-f^{11}_1\right)d_2w_2-
f^{11}_1\left(w_1\right)^2
\right)\mathcal{G}_{22}^{(1)}
,\\
&
\mathcal{G}_{44}^{(2)}=-f^{11}_1\,\mathcal{G}_{22}^{(1)}-f^{11}_2\,(w_2)^2+d_4\,w_1
,\\
&\mathcal{G}_{55}^{(2)}=\left(1-f^{11}_2\right)^2\left(\mathcal{G}_{22}^{(1)}\right)^2
-f^{11}_1\left(d_1\right)^2\,\mathcal{G}_{11}^{(1)}
-\left(
2f^{11}_1\left(1-f^{11}_1\right)d_1w_2+f^{12}_1d_2w_2
\right)\mathcal{G}_{12}^{(1)}
\nonumber\\
&\qquad\quad+
\left(
d_2w_1
-f^{11}_1\left(1-f^{11}_1\right)^2\left(w_2\right)^2
\right)\mathcal{G}_{22}^{(1)}
,
\end{align}
\begin{align}
&\mathcal{G}_{12}^{(2)}=-\left(
f^{11}_1\left(1-f^{11}_1\right)w_1+f^{11}_2w_1
\right)\mathcal{G}_{11}^{(1)}
+\left(d_3
-f^{11}_1d_1
\right)\mathcal{G}_{12}^{(1)}
,\\
&\mathcal{G}_{13}^{(2)}
=
-\left(
f^{11}_1\left(1-f^{11}_1\right)w_1+f^{11}_2w_1
\right)\mathcal{G}_{12}^{(1)}
+\left(d_3
-f^{11}_1d_1
\right)\mathcal{G}_{22}^{(1)}
,\\
&\mathcal{G}_{14}^{(2)}=-f^{12}_1\,\mathcal{G}_{12}^{(1)}-f^{12}_2w_1w_2+d_6
,\\
&\mathcal{G}_{15}^{(2)}=-f^{12}_1w_2\,\mathcal{G}_{12}^{(1)}
+\left(
\left(f^{12}_1\right)^2w_1
-f^{12}_2w_1\right)
\mathcal{G}_{22}^{(1)}
,\\
&
\mathcal{G}_{23}^{(2)}=\left(1-f^{11}_2\right)\left(
-f^{11}_1\left(w_1\right)^2+d_2w_2
\right)\,\mathcal{G}_{12}^{(1)}
+\left(
f^{11}_1\left(1-f^{11}_1\right)f^{12}_1w_1w_2+d_1d_2
\right)\mathcal{G}_{11}^{(1)}
\nonumber\\
&\qquad\quad
+
\left(
f^{11}_1f^{12}_1d_1w_2
+
\left(1-f^{11}_1\right)\left(
d_2w_2
-f^{11}_1\left(w_1\right)^2
\right)
\right)\,\mathcal{G}_{12}^{(1)}
-f^{11}_1d_1w_1\,\mathcal{G}_{22}^{(1)}
,\\
&\mathcal{G}_{24}^{(2)}=-f^{12}_2w_2\,\mathcal{G}_{11}^{(1)}
-f^{12}_1\,(
1-f^{11}_1
)w_1\,\mathcal{G}_{12}^{(1)}
-f^{12}_1d_1\,\mathcal{G}_{22}^{(1)}
,\\
&
\mathcal{G}_{25}^{(2)}=\left(\mathcal{G}_{12}^{(1)}\right)^2
-f^{12}_2\mathcal{G}_{11}^{(1)}\mathcal{G}_{22}^{(1)}
-f^{12}_1\left(1-f^{11}_1\right)w_1d_1\,\mathcal{G}_{11}^{(1)}
\nonumber\\
&\qquad\quad
+
\left(
d_3-
f^{12}_1\left(1-f^{11}_1\right)^2w_1w_2-
f^{12}_1\left(d_1\right)^2
\right)\mathcal{G}_{12}^{(1)}
-
f^{12}_1\left(1-f^{11}_1\right)d_1w_2
\mathcal{G}_{22}^{(1)}
,\\
&\mathcal{G}_{34}^{(2)}=
\left(
\left(f^{12}_1\right)^2w_2
-f^{12}_2w_2\right)
\mathcal{G}_{12}^{(1)}
-f^{12}_1w_1\,\mathcal{G}_{22}^{(1)}
,\\
&\mathcal{G}_{35}^{(2)}=
\left(f^{12}_1\right)^2w_2d_1
\mathcal{G}_{11}^{(1)}
+\left(
\left(1-f^{12}_1\right)\mathcal{G}_{22}^{(1)}
+f^{12}_1\left(f^{12}_1
\left(1-f^{11}_1\right)\left(w_2\right)^2
-w_1d_1
\right)
\right)
\mathcal{G}_{12}^{(1)}
\nonumber\\
&\qquad\quad
+
\left(
d_3-
f^{12}_1\left(1-f^{11}_1\right)w_1w_2
\right)\mathcal{G}_{22}^{(1)}
,\\
&\mathcal{G}_{45}^{(2)}=
\left(d_3
-f^{11}_1d_1
\right)\mathcal{G}_{12}^{(1)}
-\left(
f^{11}_1\left(1-f^{11}_1\right)w_2+f^{11}_2w_2
\right)\mathcal{G}_{22}^{(1)}.
\end{align}
Despite this seeming complexity,
the determinant of this matrix,
that is the Kac determinant at level two,
takes the following factorized form as expected from the representation theory
\begin{align}
&\det \,\mathcal{G}^{(2)}\left(\,Q_1,Q_2,Q_3=(Q_1Q_2)^{-1}\,\right)
=
\frac{\left(1-q\right)^8\left(1-t\right)^8\left(1+q\right)^2\left(1+t\right)^2}
{(tq)^{15}(Q_1Q_2)^{8}
(t^2+tq+q^2)(t^2-tq+q^2)}
\nonumber\\
&\times
\left(qQ_2-tQ_1\right)^2\left(qQ_2-t^2Q_1\right)\left(q^2Q_2-tQ_1\right)
\left(tQ_2-qQ_1\right)^2\left(tQ_2-q^2Q_1\right)\left(t^2Q_2-qQ_1\right)
\nonumber\\
&\times
\left(qQ_1^2Q_2-t\right)^2\left(qQ_1^2Q_2-t^2\right)\left(q^2Q_1^2Q_2-t\right)
\left(tQ_1^2Q_2-q\right)^2\left(tQ_1^2Q_2-q^2\right)\left(t^2Q_1^2Q_2-q\right)
\nonumber\\
&\times
\left(qQ_1Q_2^2-t\right)^2\left(qQ_1Q_2^2-t^2\right)\left(q^2Q_1Q_2^2-t\right)
\left(tQ_1Q_2^2-q\right)^2\left(tQ_1Q_2^2-q^2\right)\left(t^2Q_1Q_2^2-q\right).\nonumber
\end{align}
This factor leads to the denominator of the two-instanton partition function of 5d $SU(3)$ gauge theories
thought the AGT correspondence.

\section{$\boldsymbol{q}$-$\boldsymbol{W_4}$ Kac-Shapovalov matrix at level-one }

The level one Kac matrix for $q$-$W_4$ algebra in our convention of the basis of the Verma module is
\begin{align}
\mathcal{G}^{(1)}\left(w_1,w_2,w_3,w_4 \right)=
\left(\begin{tabular}{ccc}  
${ \langle{\boldsymbol{w}}\vert W^1_{1}W^1_{-1}\vert{\boldsymbol{w}}\rangle}$
&
${ \langle{\boldsymbol{w}}\vert W^1_{1}W^2_{-1}\vert{\boldsymbol{w}}\rangle}$
&
${ \langle{\boldsymbol{w}}\vert W^1_{1}W^3_{-1}\vert{\boldsymbol{w}}\rangle}$
\\
${ \langle{\boldsymbol{w}}\vert W^2_{1}W^1_{-1}\vert{\boldsymbol{w}}\rangle}$
&
${ \langle{\boldsymbol{w}}\vert W^2_{1}W^2_{-1}\vert{\boldsymbol{w}}\rangle}$
&
${ \langle{\boldsymbol{w}}\vert W^2_{1}W^3_{-1}\vert{\boldsymbol{w}}\rangle}$
\\
${ \langle{\boldsymbol{w}}\vert W^3_{1}W^1_{-1}\vert{\boldsymbol{w}}\rangle}$
&
${ \langle{\boldsymbol{w}}\vert W^3_{1}W^2_{-1}\vert{\boldsymbol{w}}\rangle}$
&
${ \langle{\boldsymbol{w}}\vert W^3_{1}W^3_{-1}\vert{\boldsymbol{w}}\rangle}$
\end{tabular}\right).
\end{align}
It is easy to compute these matrix elements by using the commutation relations.
We then obtain the following expression
\begin{align}
&\mathcal{G}^{(1)}_{11}=-f^{11}_1\left( w_1\right)^2+d_{2}w_2,\\
&\mathcal{G}^{(1)}_{22}=-f^{22}_1\left( w_2\right)^2+d_{2}w_1w_3
+d_4\frac{(1-qp)(1-t^{-1}p)}{(1-p)(1-p^2)}+c\,d_2\frac{1+p^2}{1-p^2}\\
&\mathcal{G}^{(1)}_{33}=-f^{33}_1\left( w_3\right)^2+d_{2}w_2,\\
&\mathcal{G}^{(1)}_{12}=-f^{12}_1 w_1w_2+d_{3}w_3,\\
&\mathcal{G}^{(1)}_{13}=-f^{13}_1 w_1w_3+d_{4},\\
&\mathcal{G}^{(1)}_{23}=-f^{23}_1 w_2w_3+d_{3}w_1.
\end{align}
See Section.2 for the definitions of $f^{\alpha\beta}$, $c$ and $d_n$.


\end{document}